\documentclass{article}
\usepackage{ltwol2e}

\usepackage{epsfig}

\newcommand{\MeVcc}   {\mbox{$ {\mathrm{MeV}}/c^2 $~}}
\newcommand{\GeVcc}   {\mbox{$ {\mathrm{GeV}}/c^2 $~}}
\newcommand{\GeVc}   {\mbox{$ {\mathrm{GeV}}/c $~}}
\newcommand{\bb}     {\mbox{$b \bar b$} }
\newcommand{\cc}     {\mbox{$c \bar c$} }

\begin{document}

\title{STUDY OF D$^{**}$ AND D$^{*'}$ PRODUCTION IN B AND C JETS,
WITH THE DELPHI DETECTOR}

\author{C. BOURDARIOS}

\address{Universit\'e de Paris Sud, Laboratoire de l'Acc\'el\'erateur
Lin\'eaire, B\^at. 200, B.P. 34, FR-91898 ORSAY CEDEX
\\E-mail: claire.bourdarios@cern.ch}   


\twocolumn[\maketitle\abstracts{ Using D$^{*+}$ mesons exclusively
reconstructed in the DELPHI detector at LEP, orbital and radial
excitations of non strange charmed mesons are studied. 
The multiplicities of the two narrow $D^0_1$ and $D^{*0}_2$
orbital excitations are measured in $Z^0 \rightarrow c \bar c$ and 
$Z^0 \rightarrow b \bar b$ decays.
Preliminary results are obtained on the production of broad D$^{**}$
states, using B meson semi-leptonic decays. 
A narrow signal of 66 $\pm$ 14 events is observed in the 
$(D^{*+} \pi^+ \pi^-)$ final state, interpreted
as the first evidence of the predicted $D^{*'}$ radial excitation. 
}]

\section{Introduction}

For mesons containing heavy and light quarks (Q$\bar q$), 
and in the limit where the 
heavy quark mass is much larger than the typical QCD scale, the spin 
$\overrightarrow{s_Q}$ of the heavy quark decouples from other 
degrees of freedom. Thus, for strong decays, the total (spin+orbital)
angular momentum $\overrightarrow{j_q} = \overrightarrow{s_q} 
+ \overrightarrow{L}$ of the light component is conserved. 
This heavy quark symmetry, together with quark potential models used for 
lower mass mesons, allows the masses and decay widths of heavy mesons to
be predicted \cite{HQET}. 

\begin{figure}[tbh]
\begin{center}
\psfig{figure=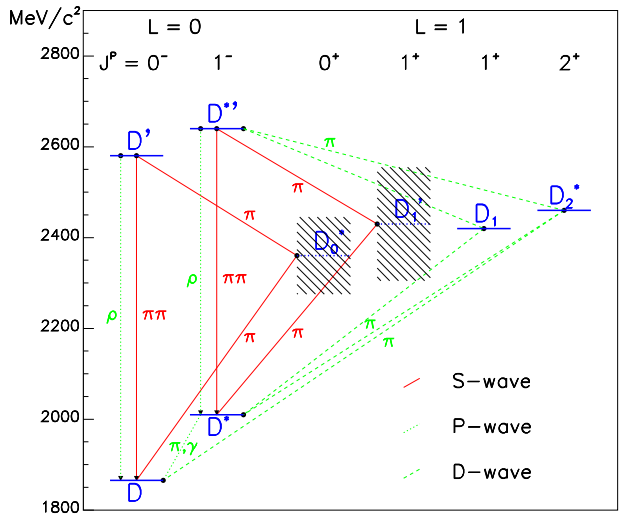,height=3.5in}
\caption{Spectroscopy of non-strange D mesons. The shaded areas show
predicted widths for these states. For clarity the expected $D_1$ and
$D^*_2$ decays involving a $\rho$ meson or $\pi\pi$ pairs are not shown.
}
\label{fig:spectro}
\end{center}
\end{figure}

The present knowledge of charmed meson spectroscopy is summarized in
Figure~\ref{fig:spectro}.
The well established D and D$^*$ mesons \cite{PDG}
correspond to the two degenerate levels of the (L=0, $j_q$ = 1/2) state. 
The two (L=1, $j_q$ =~3/2)\linebreak states have been clearly observed~\cite{PDG}, 
because they have narrow
decay widths of about 20~{\rm MeV/}$c^2$. The measured masses 
of the $D^0_1$(2420) and $D^{*0}_2$(2460) agree within 20~\MeVcc with 
the prediction of the models. 
Section 3 presents a measurement of their production rate in
\cc and \bb jets.

The (L=1, $j_q$ = 1/2) states decay through a S wave and are expected to 
have large decay widths. Up to now, they have not been observed directly,
but their total production rate is measured using B meson semi-leptonic 
decays (section 4).

In addition to these orbital excitations, radial excitations of heavy 
mesons are foreseen. The D$^{'}$ and D$^{*'}$ are expected to have 
masses of 2.58~\GeVcc and 2.64~\GeVcc respectively, 
with a 10-25~\MeVcc uncertainty on the mass predictions \cite{thdstar}. 
They are expected to decay, in S wave, into $D^{(*)}\pi\pi$.
Section 5 presents the first evidence for the D$^{*'}$ meson, 
observed in the decay mode ($D^* \pi \pi$). 

\vspace*{-1.8pt}   
\section{D$^{**}$ and D$^{*'}$ reconstruction}

DELPHI~\cite{delphi} is a multipurpose LEP detector, with special 
emphasis on precise vertex
and charged tracks momentum reconstruction, and particle identification. 
The micro-vertex detector provides 3 R$\phi$ and 2 Z hits per track,
with intrinsic resolutions of 7.6 and 9 $\mu$m. For muons of 45 \GeVc momentum,
a resolution of $\sigma(p)/p$ of $\pm$ 3\% is obtained, and the
precision of the track extrapolation to the beam collision point is
26 $\pm$ 2 $\mu$m. Kaon and pion
identification is performed using a Ring Imaging CHerenkov detector, and
the ionisation loss in the TPC, which is the main tracking device.
A total
of 3.4 million hadronic events is obtained from the 1992-1995 data,
at center-of-mass energies close to the Z$^0$ mass.

\subsection{D$^*$ reconstruction}

\begin{figure}[tbh]
\begin{center}
\psfig{figure=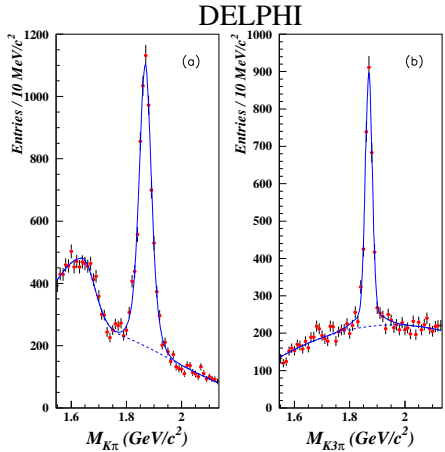,height=2.5in}
\caption{$K\pi$ and $K3\pi$ invariant mass distributions in (a) the
$D^{*+} \rightarrow (K^-\pi^+)\pi^+_*$ and (b) the $D^{*+} \rightarrow 
(K^-\pi^+\pi^-\pi^+)\pi^+_*$ decay channels. 
}
\label{fig:d0}
\end{center}
\end{figure}

All the decay channels considered here involve the $D^{*+} 
\rightarrow D^0 \pi^+_*$ decay, followed by $D^0 \rightarrow (K^-\pi^+)$
or $D^0 \rightarrow (K^-\pi^+\pi^-\pi^+)$.
\footnote{Throughout this paper, charge-conjugate states are always 
implied, and the $\pi$ from D$^*$ decay is denoted as $\pi_*$. }

To reconstruct the $D^0$ decay final state, all ($K^-\pi^+$) and 
($K^-\pi^+\pi^-\pi^+$) combinations are tried to fit a secondary 
vertex in space. Kinematical and track selection cuts are described in
detail in~\cite{n483}. 
Kaon candidates are considered if they have a momentum larger
than 1~\GeVc and, in the $K 3 \pi$ channel, a loose kaon identification
is required. The D$^0$ momentum and invariant mass are computed from the 
momenta of the decay products.
Then, all charged particles with momentum between 0.4~\GeVc and 4.5~\GeVc
and charge opposite to that of the kaon candidate are used as pion 
candidates for the $D^{*+} \rightarrow D^0 \pi^+_*$ decay. 
In the $K \pi$ ($K 3 \pi$) channel, events are selected if the mass
difference $(M_{K \pi \pi_*} - M_{K \pi}$) 
(resp. $(M_{K3\pi\pi_*} - M_{K3\pi})$) is
within $\pm$ 2 ($\pm$ 1 )~\MeVcc of the nominal value ($M_{D^*}-M_{D^0}$).
The D$^*$ candidates must have an energy fraction 
$X_E(D^*) = E(D^*)/E_{beam}$ greater than 0.25. Figure~\ref{fig:d0} 
shows the distribution of the M($K\pi$) and M($K3\pi$) invariant 
masses for the selected events. The fitted $D^0$ masses and widths are 
1868 $\pm$ 1 (1869 $\pm$ 1) and 19 $\pm$ 1 (12 $\pm$ 2)~\MeVcc.
The reconstructed D$^0$ mass is required to lie within $\pm$ 40
($\pm$ 30) \MeVcc of the nominal D$^0$ mass: 
4661 $\pm$ 88 (2164 $\pm$ 65) D$^*$ candidates are selected in the 
$K \pi$ ( $K 3 \pi$) channels. The selection efficiency is estimated,
using the simulation, to be 21\% (8\%).

\subsection{$D^0_1$, $D^{*0}_2$ and D$^{*'}$ reconstuction}

Similar selection criteria and vertex reconstruction are used to 
reconstruct narrow orbitally and radially excited states.

In the case of $D^0_1$ and $D^{*0}_2$ decaying into $D^{*+}\pi^-$,
a pion with a charge opposite that of the D$^{*+}$ is added,
and the $D^0 \pi^+_* \pi^-$ vertex is fitted.
All combinations are tried, provided the pion candidate has a
momentum larger than 1.0 (1.5)~\GeVc in the $K\pi$ ($K3\pi$) 
channel. The reconstruction 
efficiency is 14\% (6\%) in the $K \pi$ ( $K 3 \pi$) channels.

In the case of D$^{*'}$ decaying into D$^{*+}\pi^+\pi^-$, all pairs of
oppositely charged pions are used to fit a $D^0 \pi^+ \pi^-$ vertex.  
The pion candidates are required to have a momentum larger than 0.6(1.0)
{\rm GeV/c}, and those compatible with a kaon according to particle 
identification are rejected. For a signal of mass 2640~{\rm MeV/}$c^2$, the
reconstruction efficiency is 4\% (2\%) in the 
$K \pi$ ( $K 3 \pi$) channels.

In both cases, the precision on the invariant mass reconstruction is 
improved by correcting for a 4~\MeVcc shift observed in the D$^0$ mass,
by using:

\begin{equation} \begin{array}{rcl}
M(D^*\pi) = M_{(D^0\pi_*\pi)} - M_{(D^0\pi_*)} + m_{D^*} \\
M(D^*\pi\pi) = M_{(D^0\pi_*\pi\pi)} - M_{(D^0\pi_*)} + m_{D^*}
\end{array} \label{eq:mass}
\end{equation}

\noindent where $m_{D^*}$ is the nominal $D^{*+}$ mass.
The simulation predicts a resolution of about 6~\MeVcc on 
the mass reconstruction, for both radial and orbital excitations. 

\subsection{Selection of \bb and \cc samples}

Due to the relatively long lifetimes of charmed and bottomed particles,
heavy flavour events are characterized by the presence of secondary
vertices. The probability $\mathcal{P}$ that all tracks detected 
in the event come 
from the primary vertex is small: for \bb events, a purity of 90\% is 
archieved, with an efficiency of 60\%, by requiring $\mathcal{P} \le 
10^{-2}$.

Charmed mesons from $Z^0 \rightarrow b \bar b$ events are distinguished from
those in \cc events by considering both their energy and lifetime
informations. Bottom quarks fragment into a B hadron, which
subsequently decays into a D$^{*+}$ meson, whereas in \cc events charmed 
mesons are directly produced in the fragmentation process. This difference 
in the hadronization leads to a smaller energy fraction of $X_E(D^*)$
for \bb events. Also, due to the b quark lifetime, the apparent flight
of the $D^0$ meson is greater than the true decay length. 
Its measured proper time distribution is larger than the 
mean B meson lifetime, 1.6~ps, compared to a true $D^0$ 
lifetime of 0.4~ps. 

By combining these variables, \bb and \cc samples are
selected, with high purities: 92 \% for \bb, 89 \% for $c\bar c$.

In the \bb sample, the combinatorial background is higher, but is 
reduced by 50\% using the kaon identification, and also by asking that the
impact parameter of the additional pion is positive, i.e. that the 
intersection of the pion and $D^*$ directions
is on the same side of the primary vertex as the $D^*$ vertex.
As a consequence, the ratio of efficiencies 
$\epsilon(D^* \pi) / \epsilon(D^*)$ is 52\% for \bb, compared to 
62\% for \cc. 

\section{Study of narrow orbital excitations}

\begin{figure}[tbh]
\begin{center}
\psfig{figure=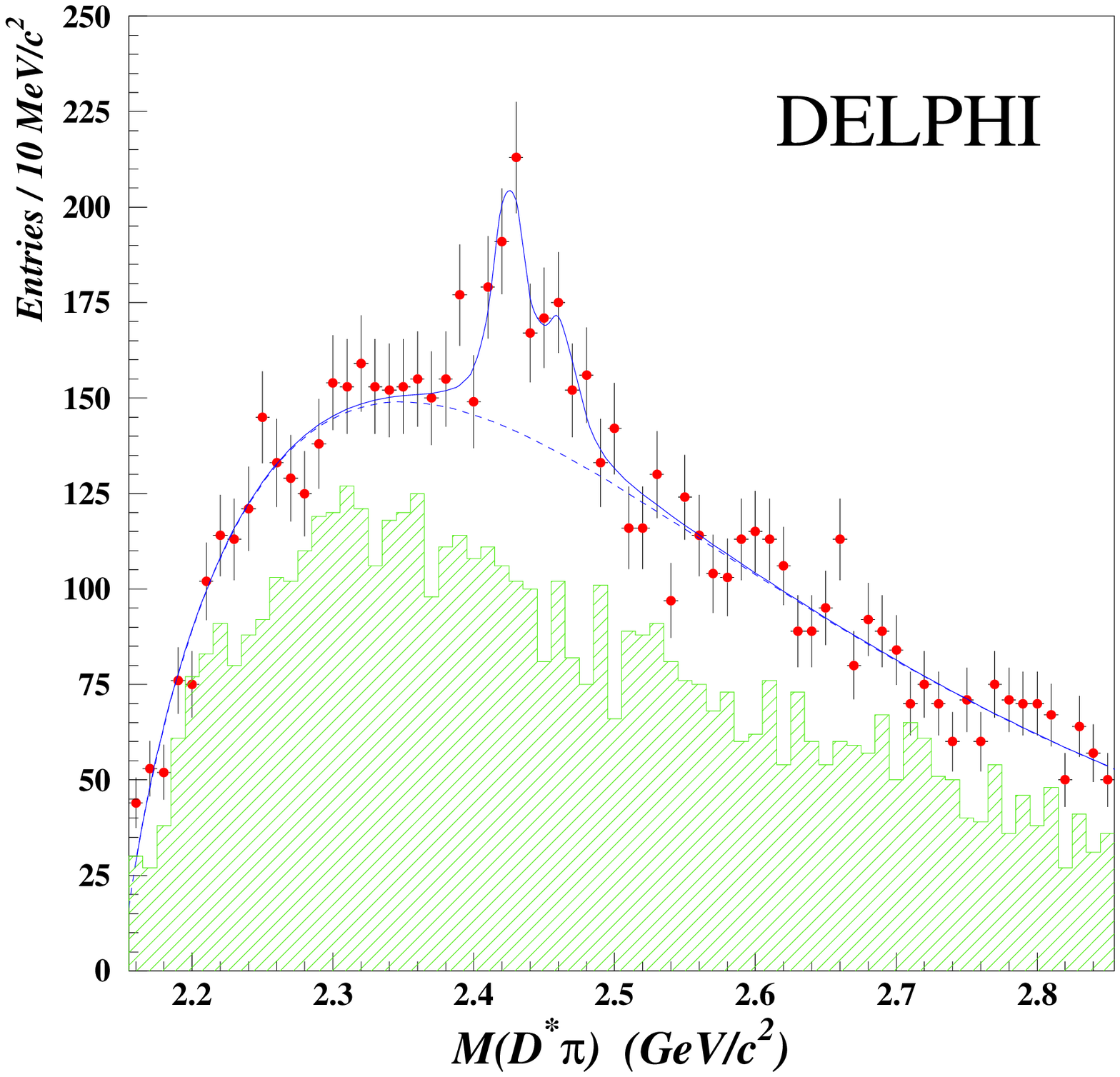,height=2.5in}
\caption{Invariant mass distributions $(D^{*+}\pi^-)$ (dots) and
$(D^{*+}\pi^+)$ (hatched histogram). The mass computation and the fit
are explained in the text. }
\label{fig:d1d2}
\end{center}
\end{figure}

Figure~\ref{fig:d1d2} shows the $M(D^*\pi)$ invariant mass distribution
obtained for the sum of the \bb and \cc samples~\cite{n240}. 
A clear excess of 
($D^{*+}\pi^-$) pairs is observed between 2.4 and 2.5 {\rm GeV/}$c^2$, corresponding 
to the two overlapping contributions of the $D^0_1$ and $D^{*0}_2$. 
They are fitted by two Breit-Wigner functions, whose widths are fixed to the
measured world average~\cite{PDG}, convoluted with the experimental 
resolution. 
A total signal of (361 $\pm$ 58) $D^0_1$ + $D^{*0}_2$ events is fitted, out
of which (65 $\pm$ 10) \% is assigned to $D^0_1$.
The masses are left free in the fit, 
and the result is $M_{D^0_1} = 2425 \pm 3 (stat)$~\MeVcc and 
$M_{D^{*0}_2} = 2461 \pm 6 (stat)$~{\rm MeV/}$c^2$, i.e. consistent with the 
world averages~\cite{PDG}.
The helicity distributions are consistent with the production of 
a $J^p=1^+$ and $J^p=2^+$ states.
 
\begin{figure}[tbh]
\begin{center}
\psfig{figure=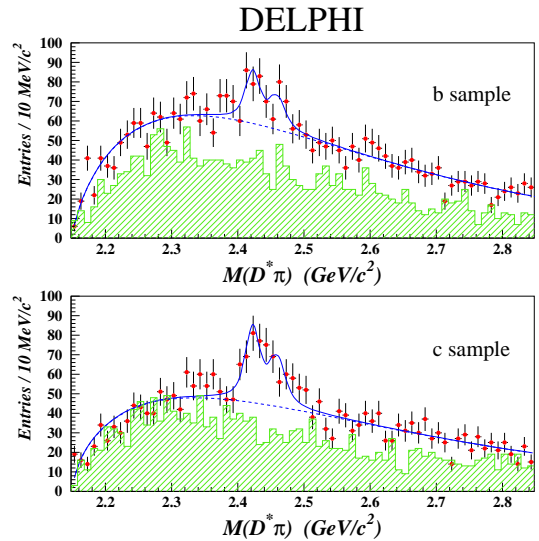,height=3.in}
\caption{Invariant mass distributions $(D^{*+}\pi^-)$ (dots) and
$(D^{*+}\pi^+)$ (hatched histogram) for the separate 
$b \bar b$ and $c \bar c$ 
samples. The mass computation and the fit are explained in the text.
}
\label{fig:ccbb}
\end{center}
\end{figure}

Figure~\ref{fig:ccbb} shows the same mass distribution, 
but for the \bb and \cc samples separately. 
The same fit is performed, but both $D^0_1$ and $D^{*0}_2$ masses and 
widths are fixed to the world average.
The result of the fit is \linebreak (97 $\pm$ 26) $D^0_1$ 
and (69 $\pm$ 27) $D^{*0}_2$ in the \bb sample,\linebreak (141 $\pm$ 26) $D^0_1$
and (104 $\pm$ 26) $D^{*0}_2$ in the \cc sample.
In order to measure the $D^0_1$ and $D^{*0}_2$ production rates,
these results are unfolded from the reconstruction
efficiencies, signal purities, 
 $D^0_1$ and $D^{*0}_2$ decay widths into $D^{*+}\pi^-$. The errors 
quoted below are
statistical only. Systematic errors are still under study, but 
smaller than the statistical errors.

The \cc sample provides direct information on the charm 
fragmentation. Results are:

\begin{equation} \begin{array}{rcl}
f(c \rightarrow D^0_1) = 1.9 \pm 0.4~(stat)~\% \\
f(c \rightarrow D^{*0}_2) = 4.3 \pm 1.3~(stat)~\%
\end{array} \label{eq:rescc}
\end{equation}

\noindent
Both results are in agreement with previous LEP and CLEO measurements.
For the $D^0_1$, the result is also in agreement with theoretical
calculations~\cite{becatini}, which predicted 1.7\%. For the $D^{*0}_2$,
the result is high compared to the expectation (2.4 \%), but has
large errors. A more precise measurement would need to use the
$D^{*0}_2 \rightarrow D^+ \pi^-$ channel, which is forbidden for the $D^0_1$.

For the \bb sample, results are:

\begin{equation} \begin{array}{rcl}
f(b \rightarrow D^0_1) = 2.2 \pm 0.6 ~(stat)~\% \\
f(b \rightarrow D^{*0}_2) = 4.8 \pm 2.0 ~(stat) ~\%
\end{array} \label{eq:resbb}
\end{equation}

\noindent
This shows that the charm fragmentation properties are similar, although
in a different environment.

\clearpage

\section{Study of broad orbital excitations in B meson semileptonic decays}

In B meson semileptonic decays, only 60\% to 70\% of the final states
are described by $D\ell\bar\nu_{\ell}$ and $D^*\ell\bar\nu_{\ell}$.
The remaining contribution is attributed to $D^{**}$.
The total production rate, including broad orbital excitations,
can be measured using the impact parameter of the pion,
denoted $\pi_{**}$, emitted in the decay chain $B \rightarrow D^{**} X 
\rightarrow (D^* \pi_{**}) X'$.

\begin{figure}[tbh]
\begin{center}
\psfig{figure=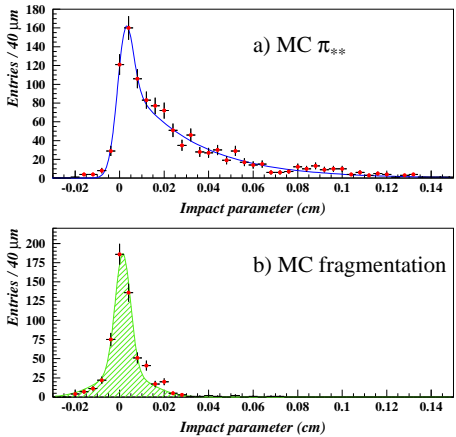,height=2.5in}
\caption{ Impact parameter relative to the primary interaction vertex 
in simulated B semileptonic decays for a) $\pi_{**}$ from $D^{**}$
decay and b) charged particles from jet fragmentation. }
\label{fig:fitmc}
\end{center}
\end{figure}

\begin{figure}[tbh]
\begin{center}
\psfig{figure=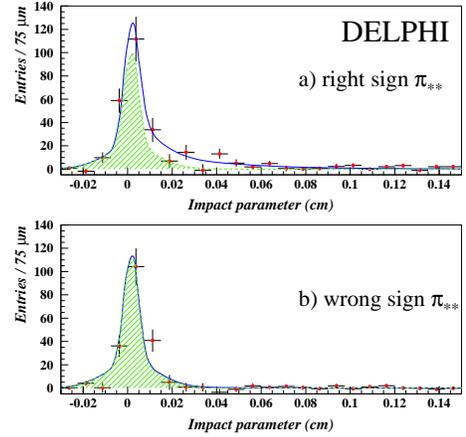,height=2.5in}
\caption{ Impact parameter relative to the primary interaction vertex
in real data for a) right charge $\pi_{**}^-$ and b) wrong charge 
$\pi_{**}^+$ candidates. The hatched area is the contribution from jet
fragmentation. }
\label{fig:fitrd}
\end{center}
\end{figure}

Events are selected if a lepton with momentum larger than 3 \GeVcc is
identified, and if its transverse momentum relative to the $D^{*+}$ is 
larger than 0.5 {\rm GeV/}$c^2$.
The kaon candidates in the $D^0$ decay must have the same charge as the
lepton. 459 $\pm$25 (288 $\pm$ 19) events are selected in 
the $K\pi$($K3\pi$) channel. 
All remaining tracks, of charge opposite to that of the $D^{*+}$, are
$\pi_{**}$ candidates. Kinematical and selection cuts are described
in~\cite{n239}. The background due to fake 
$D^{*+}$ associated to a true lepton $\ell^-$ is subtracted by using
events in the tail of the $D^{*}$ invariant mass distribution.
The contribution of true $D^{*+}$ associated to a fake lepton is subtracted 
using $D^{*} \ell$ pairs with the wrong sign combination.
The remaining 111 $\pm$ 16 events are due to true b semileptonic 
decays into $D^{*+}\ell^-X$ final state, associated with a $\pi_{**}$ 
candidate either from $D^{**}$ decay, or from jet fragmentation.
The shape of the two contributions are shown in figure~\ref{fig:fitmc}.
They are used to fit the $\pi_{**}$ impact parameter distribution
shown in figure~\ref{fig:fitrd}. From the result of the fit, the following
branching ratio is obtained:

\begin{equation} \begin{array}{rcl}
BR(B^- \rightarrow (D^{*+} \pi^-) \ell^- \bar \nu X ) \\
= 1.15 \pm 0.17 (stat) \pm 0.14 (syst) ~\%
\end{array} \label{eq:resd2}
\end{equation} 

This result significantly improves a previous \linebreak DELPHI measurement,
and is in agreement with other LEP measurements.

The ($D^{*+}\pi_{**}$) invariant mass is also reconstructed and
used to fit, in the way described in the previous section,
the $D^0_1$ and $D^{*0}_2$ narrow resonances. A signal of 26.7 $\pm$ 8.2 
$D^0_1$ is fitted, and the corresponding production rate is:

\begin{equation} \begin{array}{rcl}
BR(B^- \rightarrow D^0_1 \ell^- \bar \nu X) \\
= 0.72 \pm 0.22 (stat) \pm 0.13 (syst) ~\% 
\end{array} \label{eq:resd1}
\end{equation}

For the $D^{*0}_2$ state, 14.8 $\pm$ 7.7 events are fitted, i.e. a
signal significance smaller than 2 $\sigma$. More data would 
be necessary to estimate the corresponding branching ratio.

\section{Evidence for a narrow radial excitation}

\begin{figure}[tbh]
\begin{center}
\psfig{figure=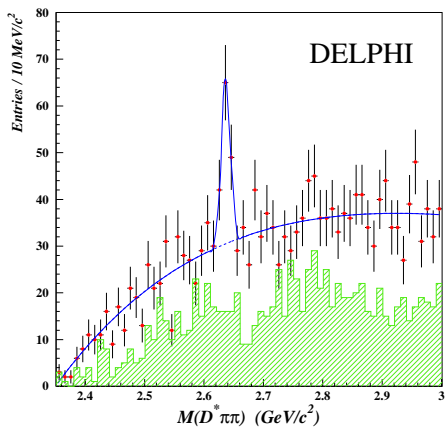,height=2.5in}
\caption{Invariant mass distributions $(D^{*+}\pi^+\pi^-)$ (dots) and
$(D^{*+}\pi^-\pi^-)$ (hatched histogram).
The mass computation and the fit are explained in the text.
}
\label{fig:dstar}
\end{center}
\end{figure}

Figure~\ref{fig:dstar} shows the invariant mass distribution obtained 
when two pions of opposite charges are added to the D$^*$ candidate, 
and using the sum of the two \bb and \cc samples. 
An excess of 66 $\pm$ 14 (stat) events is observed in the 
$(D^{*+}\pi^+\pi^-)$ combination.
The signal is fitted by a Gaussian distribution of free parameters: the 
$\chi^2$ per degree of freedom is 60/59, and would be 91/62 if 
the Gaussian was removed.
About (57 $\pm$ 10)\% of the signal is selected in the \cc sample.
 
The fitted mass is 2637 $\pm$ 2 (stat) $\pm$ 6 (syst)~{\rm MeV/}$c^2$.
It is thus consistent with the predictions for the D$^{*'}$ radial
excitation~\cite{thdstar}: 2640~{\rm MeV/}$c^2$.
Other L=2 states are predicted, but with masses higher by 
at least 50~\MeVcc.
\newpage
The width of the fitted Gaussian is 7 $\pm$ 2~{\rm MeV/}$c^2$, i.e. compatible 
with the detector resolution. Therefore, only an upper limit is
derived: the full width of the signal is smaller than 15~\MeVcc at 95 \%
C.L. There is no natural explanation of such a small value, 
neither for the D$^{*'}$ nor for higher orbital excitations~\cite{pene}.

Various checks were performed. Varying the background shape
and the kinematical cuts has no effect within statistics. No peculiar
double counting was noticed, and the signal is stable when the $\pi^*$
is added to the $D^0 \pi^+\pi^-$ tracks in the vertex fit. 
As explained above, 
mass shifts are studied using $D^0_1$ and $D^{*0}_2$ narrow states,
and a conservative
systematic error of 6~\MeVcc is attached to the mass measurement.

The production rate of this signal can be compared with that of the $D^0_1$ 
and $D^{*0}_2$ narrow states:

\begin{equation} \begin{array}{rcl}
\frac
{ < N_{D^{*'}} > \times Br ( D^{*'} \rightarrow D^* \pi^+ \pi^- ) }
{\sum_{J=1,2}
< N_{D^{(*)}_J} > \times Br ( D^{(*)}_J \rightarrow D^* \pi ) } 
\\

= 0.49 \pm 0.18 (stat) \pm 0.10 (syst)
\end{array} 
\label{eq:ratio}
\end{equation}

Most of the systematic uncertainties cancel in this ratio. The quoted 
systematics is due to the Monte-Carlo statistics, and to the uncertainties
on widths and on the kaon rejection. 
This result is compatible,
within its large errors, with the value obtained using the thermodynamical
models already mentioned for orbital states~\cite{n483,becatini}.

\section{Conclusion} 

Using about 7000 exclusively reconstructed $D^*$ mesons, 
the $D^0_1$ and $D^{*0}_2$ multiplicities are measured in \cc events, 
and found to be consistent with theoretical calculations. 
The measured multiplicities in \bb events are consistent with the
ones in \cc events, both for the $D^0_1$ and $D^{*0}_2$ .

The total D$^{**}$ production rate, involving a D$^{*+}$ in the final
state, is measured in B meson semileptonic decays. 

A narrow signal is observed in the $(D^{*+}\pi^+\pi^-)$ final state,
at the mass M = 2637 $\pm$ 2 (stat) $\pm$ 6 (syst)~{\rm MeV/}$c^2$, interpreted
as the first evidence of the predicted D$^{*'}$ meson. 

\section*{Acknowledgements}

I am very grateful to D. Bloch and P. Roudeau for their help while 
preparing this talk, and to the DELPHI collaboration for choosing
me to give it.

\end{document}